\documentclass{sig-alternate-10pt}

\usepackage{url}
\usepackage{multirow}
\usepackage{multicol}
\usepackage{graphicx}
\usepackage{graphics}
\usepackage{color}
\usepackage{subfigure}
\usepackage{amsmath}
\usepackage{nomencl}

\begin{document}

\title{Analysis of bandwidth measurement methodologies over WLAN systems}

%
%
%
%
%

\numberofauthors{2} 
%
\author{
%
%
\alignauthor
Marc Portoles-Comeras\\
       \affaddr{Centre Tecnològic de Telecomunicacions de Catalunya (CTTC)}\\
       \affaddr{Castelldefels (Barcelona), Spain}\\
       \email{mportoles@ctt.cat}
\alignauthor
Albert Cabellos-Aparicio\\
       \affaddr{Universitat Politècnica de Catalunya}\\
       \affaddr{Barcelona,Spain}\\
       \email{acabello@ac.upc.edu}
\and
\alignauthor Josep Mangues-Bafalluy\\
       \affaddr{Centre Tecnològic de Telecomunicacions de Catalunya (CTTC)}\\
       \affaddr{Castelldefels (Barcelona), Spain}\\
       \email{jmangues@ctt.cat}
\alignauthor Jordi Domingo-Pascual\\
       \affaddr{Universitat Politècnica de Catalunya}\\
       \affaddr{Barcelona,Spain}\\
       \email{acabello@ac.upc.edu}
}

\maketitle

\begin{abstract}
WLAN devices have become a fundamental component of nowadays network deployments. However, even though traditional networking applications run mostly unchanged over wireless links, the actual interaction between these applications and the dynamics of wireless transmissions is not yet fully understood. An important example of such applications are bandwidth estimation tools. This area has become a mature research topic with well-developed results. Unfortunately recent studies have shown that the application of these results to WLAN links is not straightforward. The main reasons for this is that the assumptions taken to develop bandwidth measurements tools do not hold any longer in the presence of wireless links (e.g. non-FIFO scheduling). This paper builds from these observations and its main goal is to analyze the interaction between probe packets and WLAN transmissions in bandwidth estimation processes. The paper proposes an analytical model that better accounts for the particularities of WLAN links. The model is validated through extensive experimentation and simulation and reveals that (1) the distribution of the delay to transmit probing packets is not the same for the whole probing sequence, this biases the measurements process and (2) existing tools and techniques point at the achievable throughput rather than the available bandwidth or the capacity, as previously assumed.
\end{abstract}


\section{Introduction}

WLAN devices have become a fundamental component of nowadays network deployments. They can be found
in scenarios that range from simple home networks to complex mesh-like multi-radio multi-hop
infrastructures. However, even though traditional networking applications run mostly unchanged over
wireless links, the actual interaction between these applications and the dynamics of wireless
transmissions is not yet fully understood.

Bandwidth measurement tools and techniques are an example of such applications. Bandwidth
measurements have become a mature research topic with well-developed results both at a practical
level (e.g. \cite{1,17,18,19,20,23,24}) and, lately, at a more fundamental
level \cite{14,14b}. However, various preliminary studies
have shown that the application of these results to WLAN environments is not straightforward
(\cite{2,3}).

The main reasons for this reside on the assumptions taken to develop bandwidth measurement models
and tools. On one side, traditional active measurement techniques are based on the concept of a
single bit-carrier multiplexing several users in FIFO order (e.g. \cite{1}). Additionally, it is commonly assumed
that communication links present a constant transmission rate along the measurement process.
Further, another common assumption to take is that the impact of low-layer overheads can be
neglected and measurements taken with a given packet size can be easily extended to packets of
different sizes.

These assumptions do not hold any longer in the presence of wireless links. First, multiple-user
access schemes such as the CSMA/CA compromise the FIFO assumption \cite{3}. Second, the service rate may change along the measurement process \cite{8}. Finally, previous studies \cite{3,5} have shown that bandwidth metrics of a wireless link cannot be easily normalized and that measurements have to take into account the packet size used.

This paper builds from these observations in order to analyze the interaction between probe packets
and WLAN transmissions in active measurement processes. Specifically, the paper proposes a model
that better accounts for the particularities of wireless links. The model is validated through
extensive experimentation and simulation and is used to derive the specific dynamics of probing
packets in the presence of WLAN links. The model is then used to obtain a complete characterization
of bandwidth measurements gathered using active probing sequences. The paper presents both fluid
and non-fluid approaches to the bandwidth measurement problem. For the non-fluid approach we extend
a recently developed mathematical framework \cite{14} to cope with the specificities of WLAN
transmissions.

The contributions of the paper are the following:
\begin{itemize}
                 \item First, it reveals how the distribution process describing the delay
                 to transmit probing packets in a WLAN system is not the same for the whole
                 probing sequence. Instead, the distribution follows a transitory regime before
                 reaching a stationary state.
                 \item Second, taking a fluid approach to the bandwidth measurement problem, it shows that traditional
                 tools point at the \emph{achievable throughput}, rather than the \emph{available bandwidth} or the
                 \emph{capacity} when applied over wireless links.
                 \item Third, it shows how dispersion based measurements based on a short number of
                 probing packets are biased measurements of the \emph{achievable throughput}. The
                 origin of this bias lies on the transitory detected in the service delay of
                 probing packets.
\end{itemize}

The results described here have several consequences that transcend the scope of the paper and can
be useful to the research community. First, we show how the packet pair technique \cite{24}, widely used in
the wireless mesh routing literature \cite{22}, constitutes a biased measure of the \emph{achievable
throughput}. We show also that it can only be used to measure the transmission rate of a wireless
link in a very particular circumstance (i.e. when there is no contending traffic). Second, we
introduce a simple yet effective method to improve the accuracy and convergence properties of
bandwidth measurement tools. Interestingly this method not only improves measurements in wireless
scenarios but also in wired ones.

The rest of the paper is organized as follows. Section 2 introduces the tools that have been used
throughout the paper in order to validate results. Section 3 introduces the proposed model for WLAN
links. The model is validated with extensive measurements. Section 4 introduces the basics of
bandwidth measurement over WLANs and presents results when taking a fluid approximation to the
problem. Section 5 relaxes the fluid approximation, adapts a recent analysis framework to the
specificities of WLAN transmissions and derives possible biases in the results of the measurement
process when using dispersion based techniques. Section 6 discusses the consequences of some of the
findings presented. Finally, section 7 concludes the paper.

\section{Validation Setup}

The study presented in this section is based on theoretical analysis, simulation and
experimentation. This section introduces the simulation and experimentation settings used to gather
measurement data and validate theoretical findings.

Experimentation has been carried out within the EXTREME framework (see \cite{10}). This is a
multi-purpose networking experimental platform. The main advantage  of this platform is its high
automation capabilities that allow automatic execution, data collection and data processing of
several repetitions of an experiment.

The WLAN devices used are Z-COM ZDC XI-626 cards which carry the popular Prism chipset. These
wireless devices are controlled using computer nodes of the EXTREME cluster. In all cases these
nodes are Pentium IV PCs with a 3GHz processor, 512MB of RAM memory and running Linux OS, with
kernel 2.4.26. To control these devices, the EXTREME automation system makes use of the wireless
extensions API.

In order to generate the traffic (probing and cross-traffic), we make use of the Multi-GENerator
toolset \cite{11}. However, in order to increase the accuracy of the time-stamping procedure, both
at sender and receiver sides, network device drivers have been conveniently modified to timestamp
packets just before they are laid down to the hardware (sending side) and just after getting them
from the hardware (receiving side). This follows some of the ideas described in \cite{12}.

\begin{figure}
\centering
\includegraphics[height=40mm]{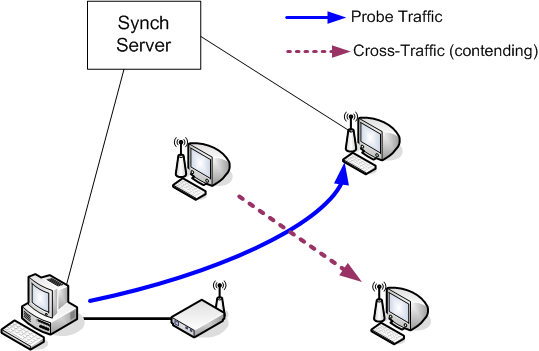}
\caption{Experimental/simulation scenario} \label{fig:1}
\end{figure}

Figure \ref{fig:1} shows the basic setup used throughout the section for experimentation. The
probing traffic is sent between two stations that are conveniently synchronized. This
synchronization is achieved by sending frequent NTP updates through a parallel wired interface
between the NTP server and the measurement nodes. Using this method we achieve accuracies of delay
measurement in the order of ten microseconds.

Unless specified the cross-traffic generated follows a Poisson distribution.

Some of the experiments required a large amount of repetitions to achieve accurate convergence of
results. Since this is difficult to achieve in a testbed we have also used a simulator.
Specifically we have replicated the tesbed (figure \ref{fig:1}) using NS2 (ver. 2.29 \cite{13}).
The main difference between the testbed and the simulator is that the latter includes scenarios
with up to 5 contending nodes. Following some recent research results \cite{25} both the testbed
and the simulator went through a thorough calibration process in order to assure that the results
gathered are comparable.

The simulator uses the NO Ad-Hoc Routing Agent. This agent supports static routing configurations over wireless networks and does not send any routing related packets. This avoids possible interferences with probe or cross-traffic. Regarding the configuration, all the experiments use the default MAC and PHY 802.11 layers included into the NS2 package. The queues used are infinite, this way we avoid
dealing with packet losses, which are irrelevant for our study. Finally all the wireless nodes are
static and equally spaced from the Access Point. The physical transmission rate is set to 11Mbps
and RTS/CTS is not used.

Finally, we have also developed a queuing simulator using Matlab. The motivation for this is that
the probing process in a WLAN presents multiple components that are difficult to isolate from each
other in an experimentation setting or even through simulations. The queuing simulator convolves a series
of packet arrivals with a series of service times in order to measure several metrics such as the
queuing length distribution and the output dispersion (inter-arrival) of packets. The input
parameters are gathered from experimentation measurements in order to keep the results as close to
the real behavior as possible.

Unless noted otherwise the results presented in this work have been obtained from repeating
experiments over 80 times while the simulations have been repeated 25.000 (NS2) to 70.000 (Matlab)
times.

\section{Model of WLANs for bandwidth measurements}
Traditionally the development of bandwidth measurement techniques rely on a series of well-accepted
assumptions. However some of them do not hold true in the presence of wireless links. This section
reviews these assumptions, proposes a more generic model accounting for WLAN specificities and
validates the model through extensive measurements.\\
\\

\begin{figure}
\centering
\includegraphics[height=30mm]{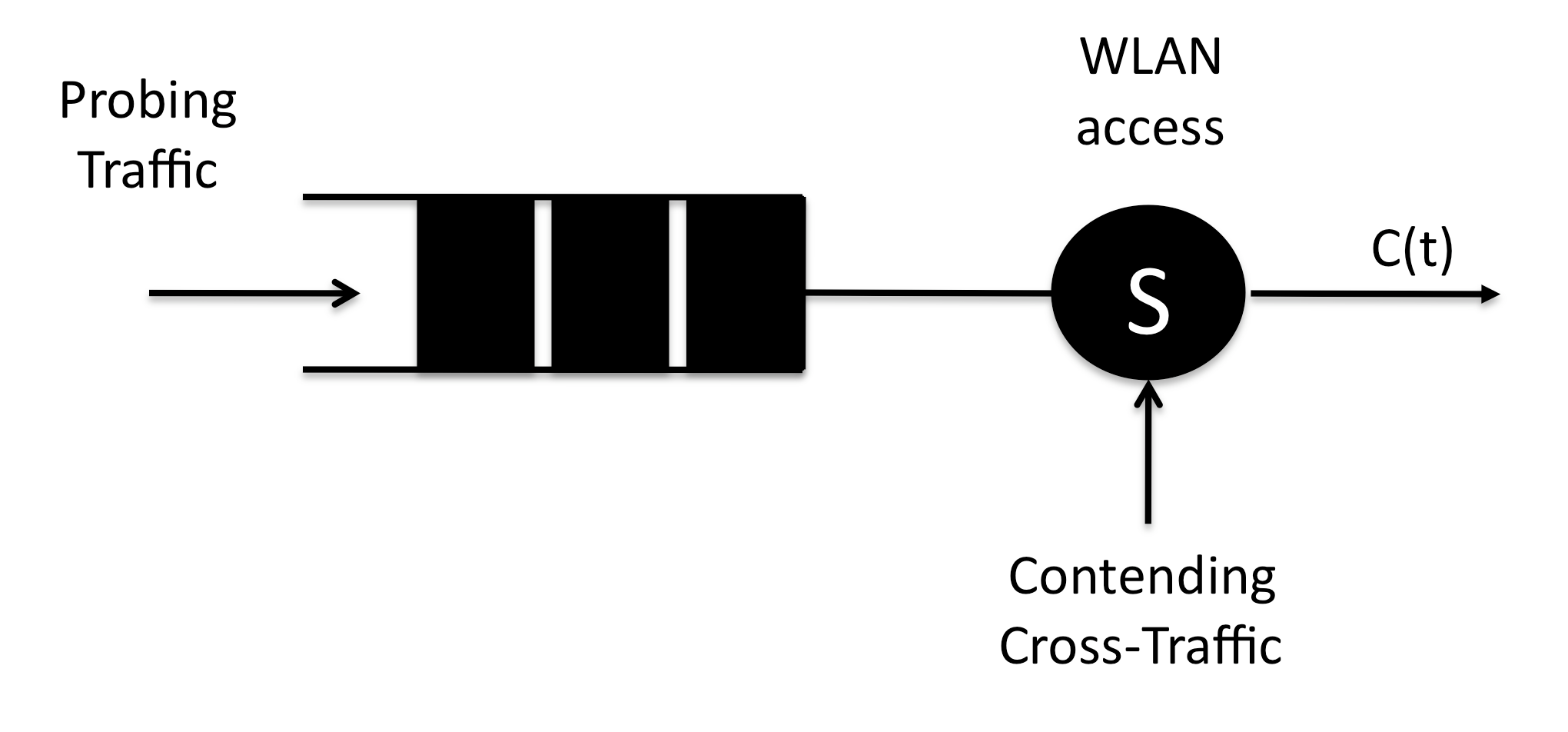}
\caption{Model of the interaction between probing traffic and (contending) cross-traffic in a WLAN
system} \label{fig:2}
\end{figure}

\subsection{Model of a WLAN link}

Developments related to bandwidth measurements usually rely on three assumptions that cannot be
taken in WLAN environments. First, a vast majority of studies consider network links as single
bit-carriers that multiplex multiple users in FIFO order \cite{1}. Second, it is commonly assumed that
communication links present a constant (fixed) transmission rate along the measurement time. Third,
the impact of low-layer overheads is usually neglected and tools are developed considering that
results gathered with packets of a given size can be extended to other sizes.

Firstly, Multiple-access schemes related to WLAN links prevent taking the FIFO assumption of
traditional models. As a result of using techniques such as the DCF mechanism in the IEEE 802.11
MAC protocol, probing traffic and cross-traffic are scheduled to use the channel in a non-FIFO
manner \cite{3}. A consequence of this is that the delay that packets experience once they are at the head
of the transmission queue until they are completely transmitted becomes a random process. This
random process depends on the amount of traffic contending for channel access at every instant and
the specific scheduling algorithm used.

Additionally, the variability of wireless channels and the protection mechanisms usually adopted
invalidate the assumption of a constant transmission rate. ARQ mechanisms or rate adaptation
strategies change the amount of time necessary to transmit a given packet in case of a variation in
channel propagation conditions. Hence breaking the assumption of a constant transmission rate \cite{8}. Again, as
a consequence, the delay to service packets at the head of the transmission queue becomes a random
process.

Finally, recent studies \cite{3,5} have shown that in
WLAN environments bandwidth measurements taken with packets of a given size cannot be easily
extended to other packet sizes. The origin of this are the large low layer overheads associated to
WLAN transmissions. This assumption was usually used to normalize both probe traffic and
cross-traffic to a common unit (i.e bits per second) and led developing measurement tools
irrespective of the packet size used. In WLANs the interaction between cross-traffic and probing traffic cannot be normalized to bits per second.
Instead it has to be studied from the perspective of time (transmission time, channel utilization time, scheduling delay, etc.).

Taking all these observations into account, figure \ref{fig:2} presents the model used all along
the paper. Probing packets enter a transmission queue and get service in FIFO order. Once at the
head of the transmission queue, they suffer a random transmission delay associated to the
scheduling mechanism and/or variations of channel propagation characteristics. As it can be noted
cross-traffic that contends for channel access is not considered from a packet or bit per second
perspective but it is included in the random service delay that probing packets suffer.

This model is non-parametric, hence it does not make any assumption about the distribution of the
service delay of probing packets.

\subsection{Analysis of service delay}

\begin{figure}
\centering
\includegraphics[height=50mm]{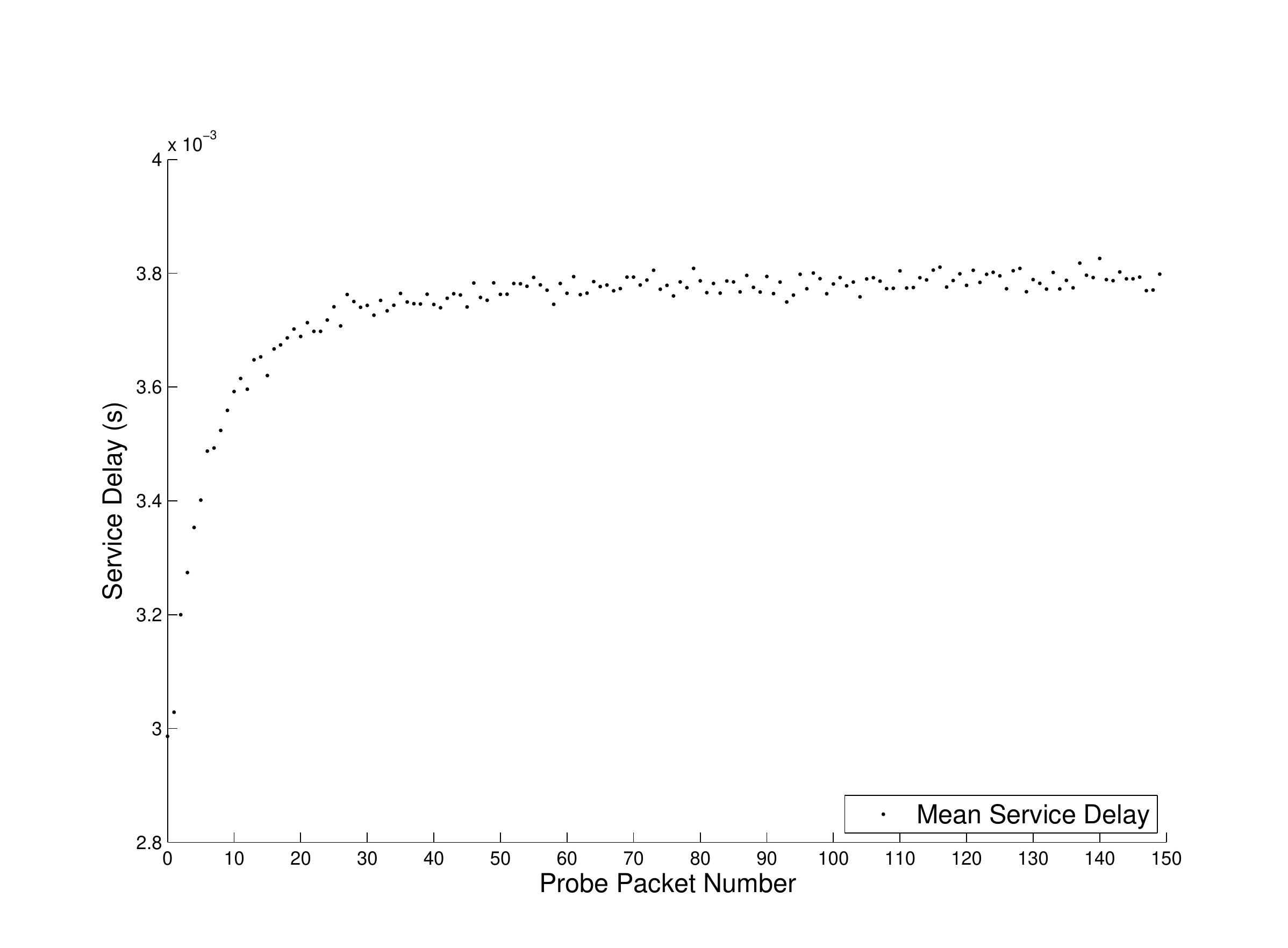}
\caption{Mean Service Delay vs. Probe packet num} \label{fig:4}
\end{figure}

This section analyses the characteristics of the service delay that probe packets experience. We
take as service delay the time since the probing packet is ready to be transmitted (it is at the
head of the transmission queue in figure \ref{fig:2}) until it is completely transmitted. The service
delay in WLANs has been repeatedly studied in the literature. Indeed, different researchers have
analyzed its exact distribution using stochastic tools, such as Markov Chains \cite{6,9},
others show how the exponential distribution provides a good fit \cite{7}. All these studies have
focused on the distribution of the service delay in \emph{stationary} state. However, in general,
bandwidth measurements are gathered using packet trains with a limited length (limited number of
packets). As a consequence, for the purpose of this work, we are interested in analyzing how the
service delay evolves over time as an increasing number of probing packets are sent through the WLAN.

In order to illustrate this evolution first consider the following experiment: using NS2 we send
1000 probe packets at a given rate (5Mbps) and with a static load of contending cross-traffic
(4Mbps). We have repeated the experiment 25000 times and, for each probe packet (indexed from 1 to
1000), we compute the distribution of the service delay (considering all the repetitions).

\begin{figure}
\centering
\includegraphics[height=50mm]{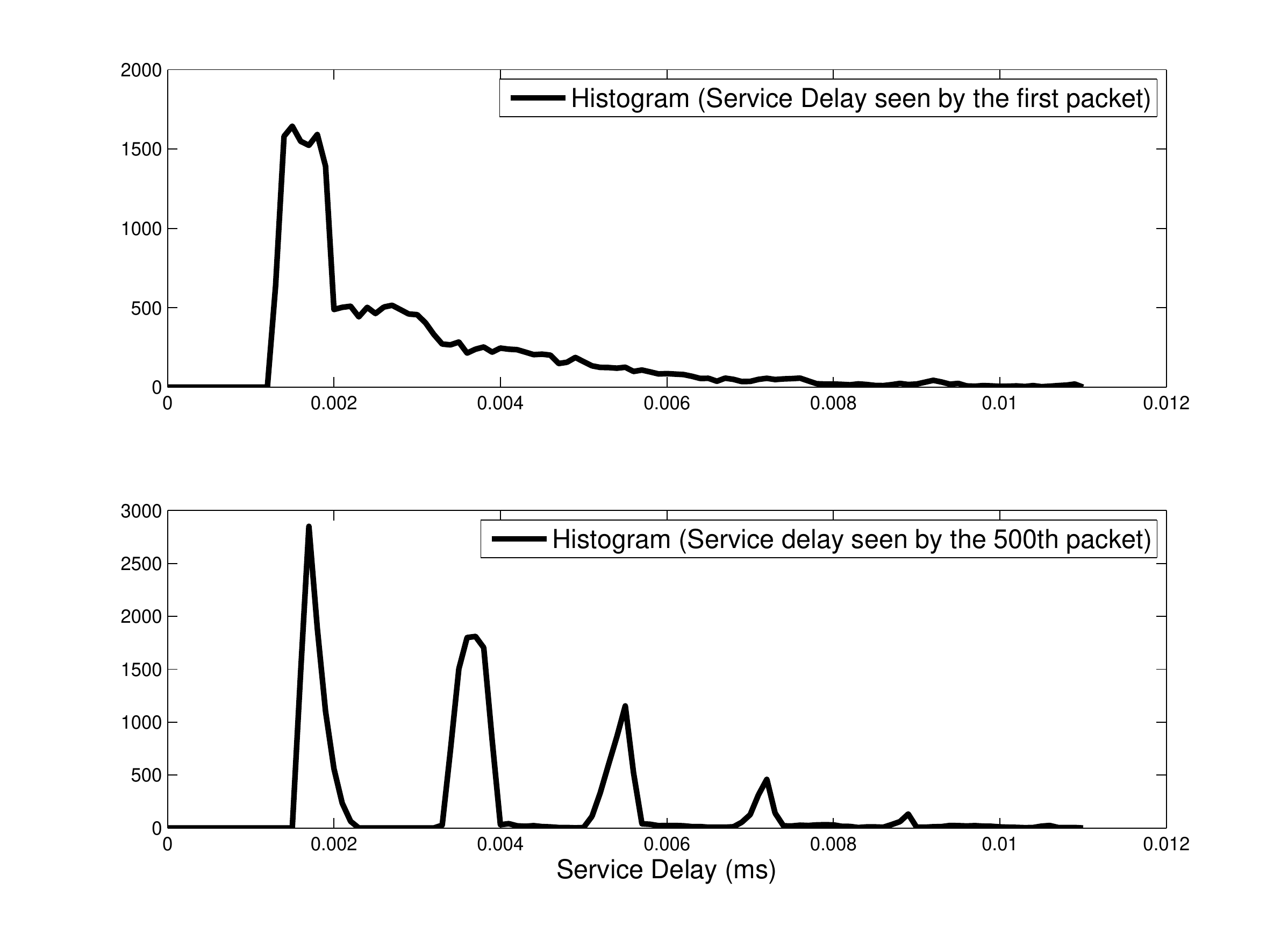}
\caption{Histogram of the s.d seen by the first and 500th packet (simulation)} \label{fig:5}
\end{figure}

Figure \ref{fig:4} plots the average service delay that each one of the first 150 packets observes.
The figure shows how the average service delay perceived by the first packets is lower than for the
rest of them. This suggests that, in fact, the distribution of the service delay changes as more
probe traffic keeps on arriving to the WLAN link. In order to verify this hypothesis, figure
\ref{fig:5} plots the histogram of the service delay as seen by the first probe packet and by the
500th. As the plot shows, the distribution changes significantly. The main rationale behind this is
that as new probing packets keep on arriving they keep on increasing the load of the network until
reaching a stationary state of interaction with the (contending) cross-traffic.

\begin{figure}
\centering
\includegraphics[height=50mm]{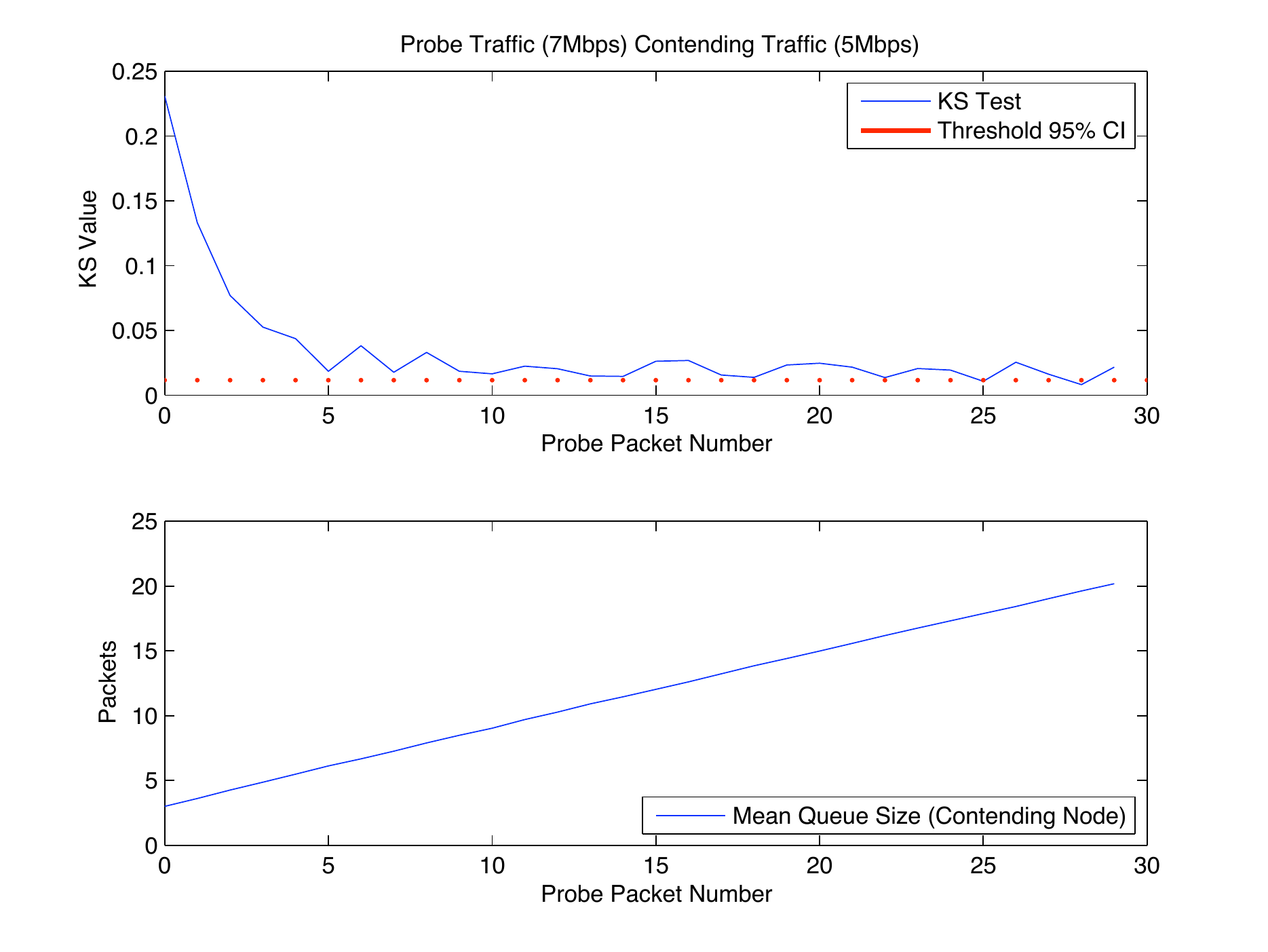}
\caption{Analysis of the distribution (7Mbps probe-traffic rate, 5Mbps cross-traffic rate (Top)
KS-Test (Bottom) Mean contending node's queue size} \label{fig:8}
\end{figure}

Let us focus again on figure \ref{fig:5} (bottom). This histogram plots the distribution of the
service delay when the link is saturated. As we can see in the figure, there are 6 visible peaks.
These peaks are related to the amount of contending rounds that a probe packet can lose before
being successfully transmitted. The first peak is the probability that the probe packets wins the
first contending round, the second peak is the probability that loses the first but wins the second
and so on. Given the example of \ref{fig:5} (bottom) we could state that, with high probability, a
probe packet will need no more than 6 contending rounds to be transmitted. This means that in order
to see the distribution figure \ref{fig:5} (bottom) the queue of the contending node must contain around 6
packets, otherwise it will not be possible for a probe packet to require 6 contention rounds to be
transmitted. This suggests that the presence and duration of this transitory depends on the queue
size of the contending node.

\begin{figure}
\centering
\includegraphics[height=50mm]{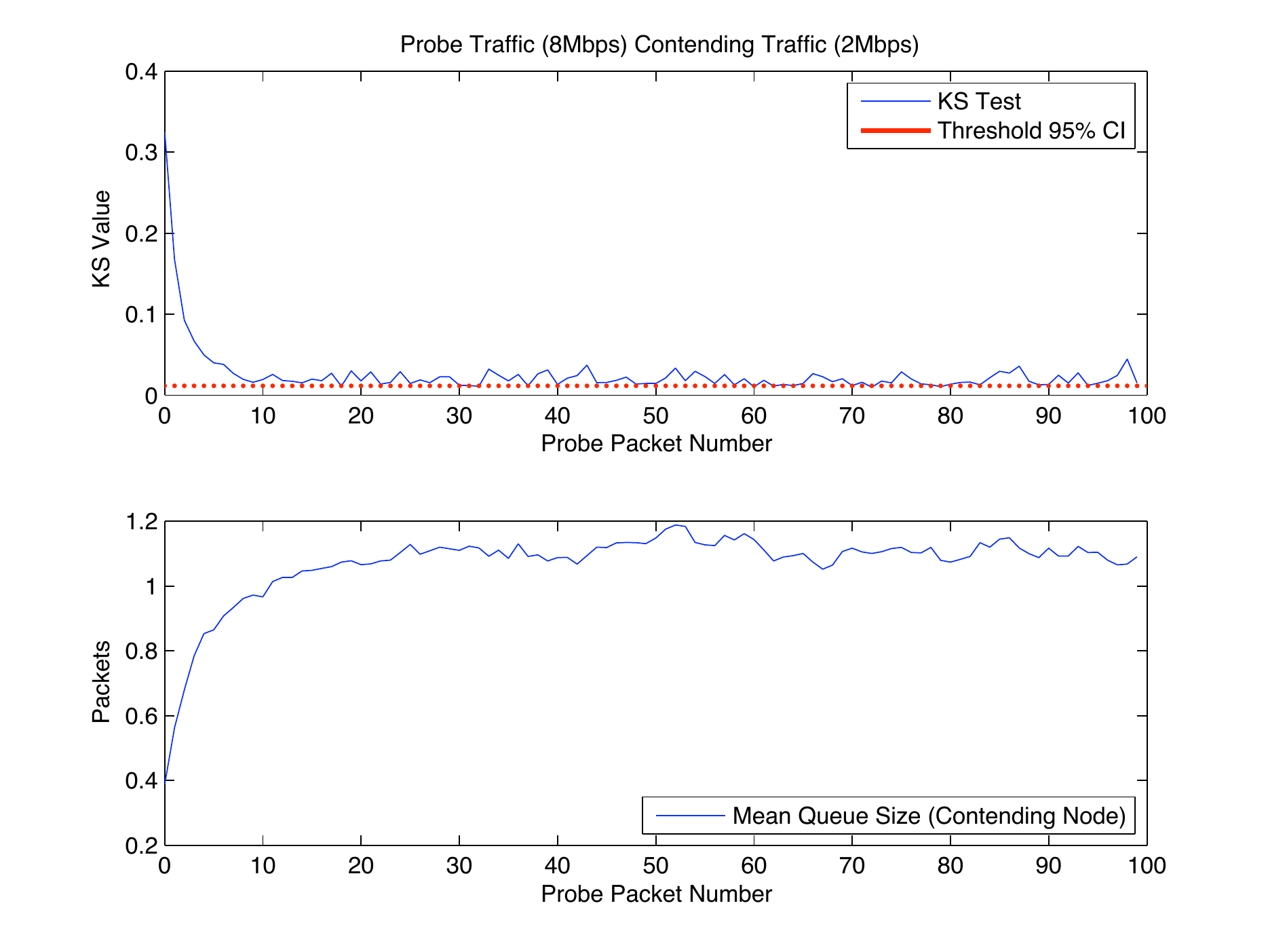}
\caption{Analysis of the distribution (8Mbps probe-traffic rate, 2Mbps cross-traffic rate (Top)
KS-Test (Bottom) Mean queue size} \label{fig:6}
\end{figure}

In order to validate this hypothesis we use the well-known Kolmogorov-Smirnov\footnote{Since
we are using the KS test to compare two empirical discrete distributions we convert one of them to
a continuous one using linear interpolation.} (KS) goodness-of-fit test \cite{15} to
compare the resemblance of the delay distribution suffered by every probing packet during the
transitory and the delay distribution once probing packets have reached a stationary state. The KS
test is non-parametric and analyzes whether two different sets come from the same random
distribution. Using this test we compare the distribution of each individual packet in the probing
sequence with the service delay distribution of the last 500 probing packets. As mentioned, our
hypothesis is that the duration of this transitory regime is related with the queue size of the
contending station. Hence we compare the result of the KS-test with the mean queue size of the
contending node. The queue size is measured when the probe packet is successfully transmitted.

 Consider the experiment depicted in figure \ref{fig:8} where both probe and cross traffic enter backlog after a certain amount of time. The top plot shows the result of the KS-test on a per probe packet basis. The bottom plot shows the corresponding average queue size at the station that is contending for channel access. It can be seen that the fifth probing packet observes a distribution of service delay that presents a close fit with the stationary distribution (the result of the KS-test approaches the red line that corresponds to the 95\% of c.i.). The bottom plot shows how when the fifth probing packet is transmitted the queue size of the contending node reaches 6 packets. This would confirm the hypothesis of the inter-relation between the queue size of the contending node and the stationarity of the service delay of probing packets.

To further validate this hypothesis we repeat the experiment but considering two cases when the contending station does not reach backlog.

\begin{figure}
\centering
\includegraphics[height=50mm]{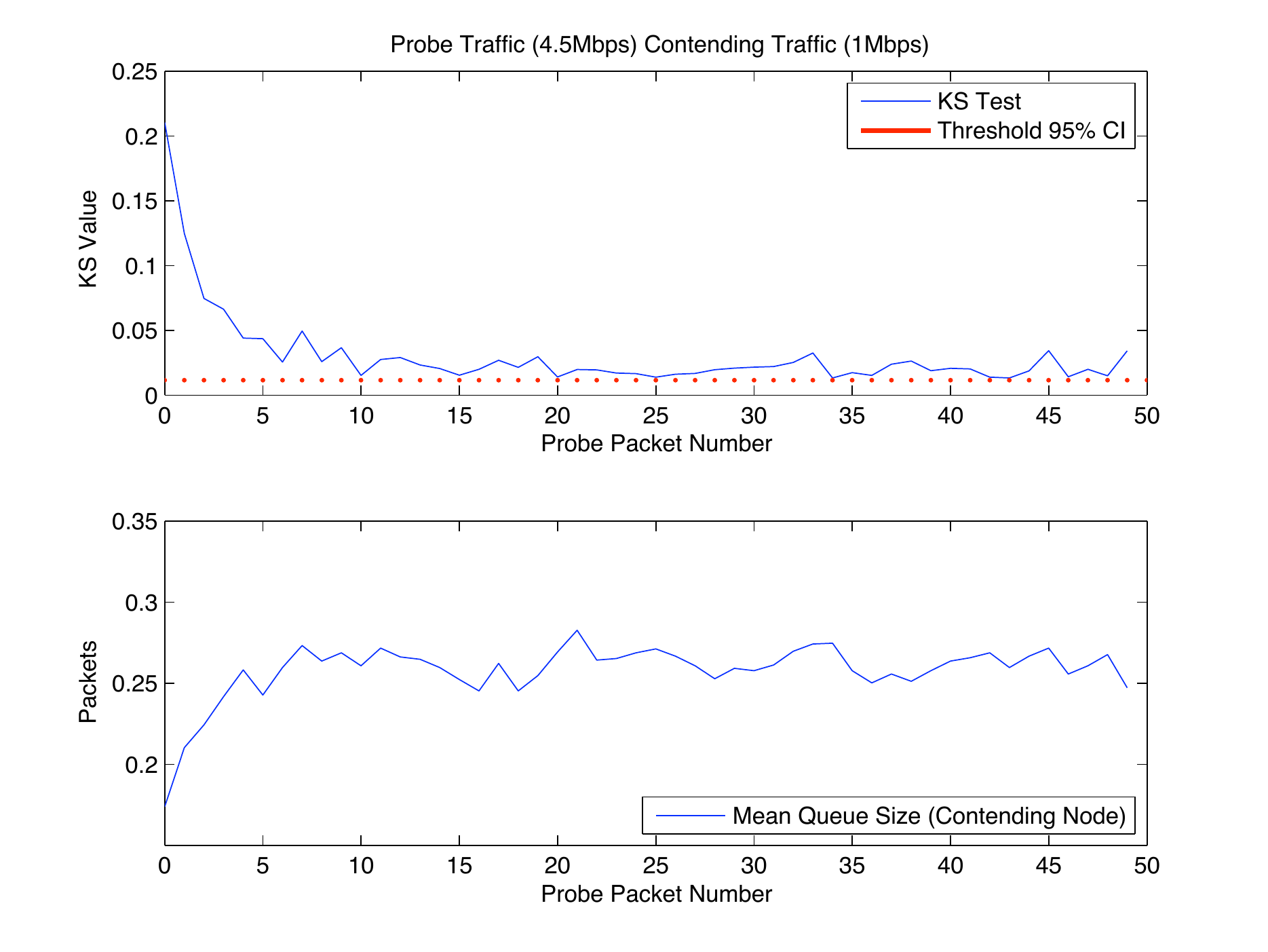}
\caption{Analysis of the distribution (4.5Mbps probe-traffic rate, 1Mbps cross-traffic rate (Top)
KS-Test (Bottom) Mean contending node's queue size} \label{fig:7}
\end{figure}

First consider the experiment depicted in figure \ref{fig:6}. The KS-test plot shows how after 10 probing packets have been transmitted the distribution of the service delay presents a close fit to stationarity (i.e. the KS-test value approaches the 95\% threshold). Observing then the queuing evolution of the contending station it can be seen that after the 10th packet is transmitted the queuing length reaches a stable average value around one. Figure \ref{fig:7} presents a similar effect. The difference in this case is that neither the probing station nor the contending node enter backlog at any time. However, the inter-relation between the transitory evolution of the service delay distribution and the number of packets in the queue of the contending node is still evident.

\begin{figure}
\centering
\includegraphics[height=30mm]{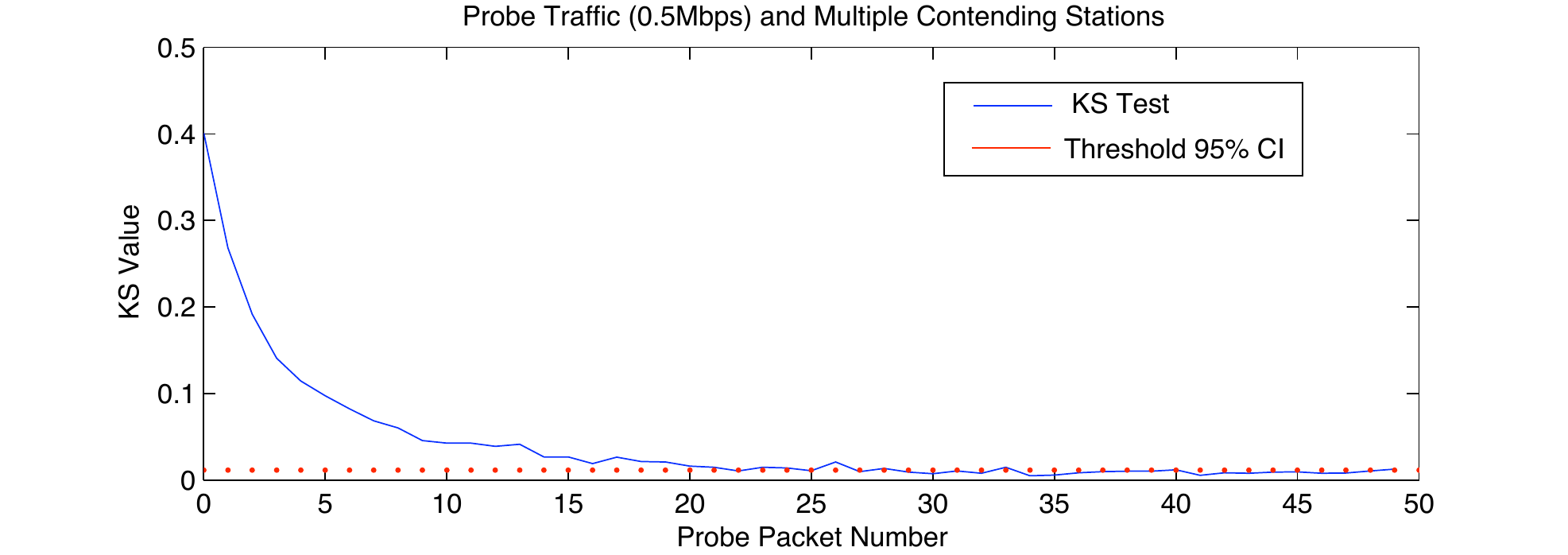}
\caption{Analysis of the distribution (complex case)} \label{fig:9}
\end{figure}

\begin{figure}
\centering
\includegraphics[height=50mm]{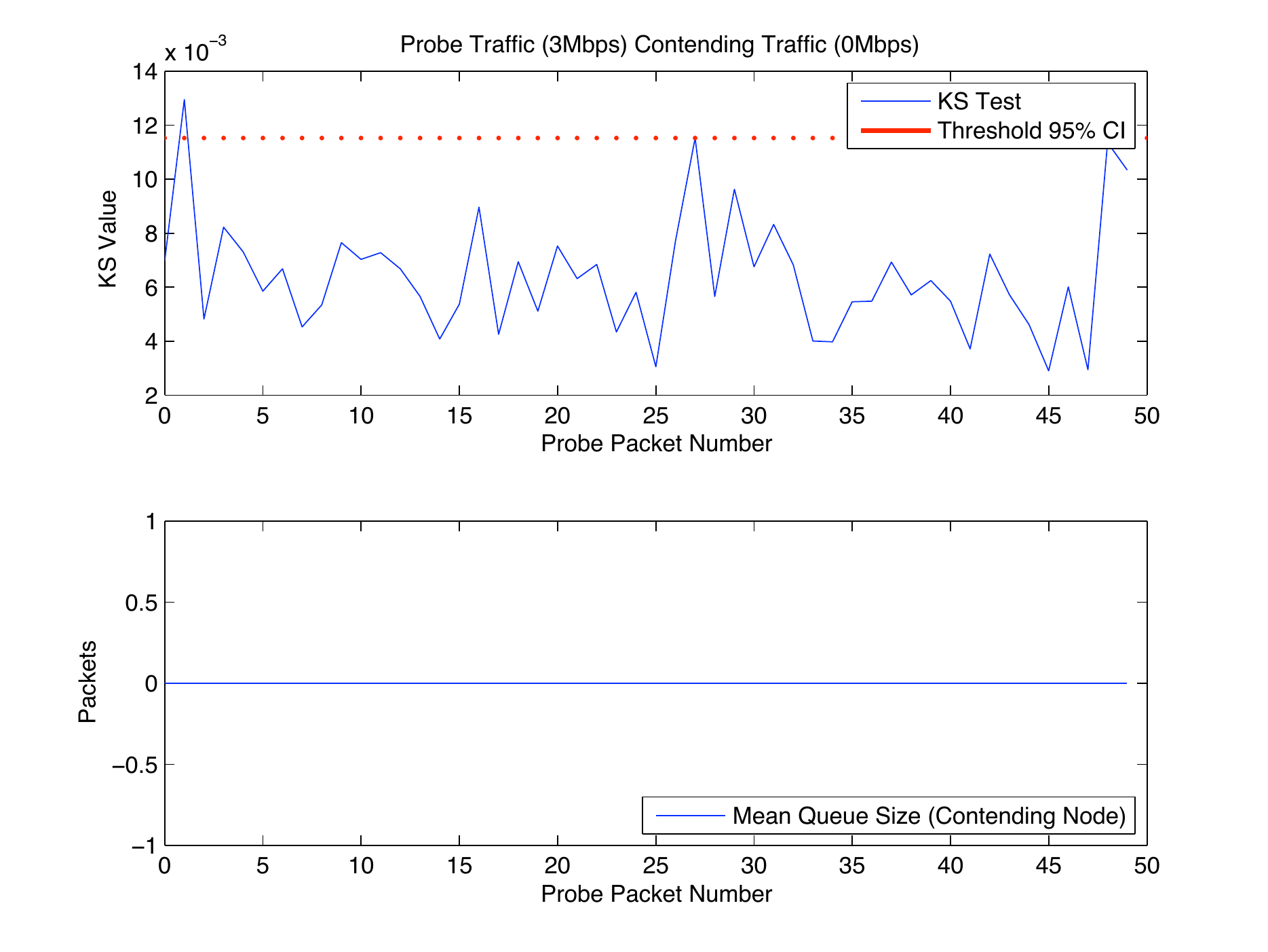}
\caption{Analysis of the distribution (3Mbps probe-traffic rate, no contending cross-traffic}
\label{fig:10}
\end{figure}

Finally we have also experimented with more complex scenarios. As an example consider figure
\ref{fig:9} that shows the KS-test for a case with 4 contending stations using different packet
sizes (40, 576, 1000 and 1500 bytes) and the following rates respectively (0.1, 0.5, 0.75 and
2Mbps). Again the figure reveals a transitory regime in the distribution of the service delay, also
related with the queue size of the contending nodes. As the figure shows we need to send tens of
packets until reaching a stationary state. We have simulated more cases with different degrees of
complexity obtaining similar results. The transitory is present in all the cases that probe traffic
has to contend for accessing the air. As a final example consider the experiment depicted in
figure \ref{fig:10} where no contending traffic is present. As the figure reveals the transitory is
non-existent.

\subsection{Consequences of the observations}

The analysis about the service delay reveals two important issues affecting the bandwidth measurement task.

On one side, the service delay of the probing sequence undergoes a transitory (in distribution)
before reaching a stationary state. This implies that the first packets of a probing sequence do
not capture the long-term behavior of larger flows and represent biased samples of the stationary
interaction between the probing flow and cross-traffic. This observation has a direct impact on
bandwidth measurement tools that generally use short trains of packets to support measurements.

On the other side, once reaching the stationary state we can see that the expectation of the
service delay remains stable no matter how many probing packets we inject in the transmission queue
(see figure \ref{fig:4}). Even more, once contending stations reach a backlog state, no matter how far we
increase the probing rate that the service delay distribution remains constant. In WLAN terms, at
this point the probing traffic has reached its fair share of the channel and it cannot get a higher
piece of it, no matter how hard it tries.

\section{Basics of Bandwidth Measurements over WLANs}

This section reviews some of the basic concepts related to bandwidth measurements in the presence
of WLAN links. The first part of the section reconsiders bandwidth metrics in the WLAN context.
Recent literature \cite{2} suggested that the same
metrics that used to be measured in wired environments do not necessarily hold in wireless
scenarios. The second part of the section presents rate response curves when placing fluid
assumptions over the model of the WLAN introduced in the previous section. This reveals an important result
regarding the identifiability of certain metrics in WLAN environments. Fluid curves are later used as a reference for the non-fluid analysis of section 5.

\subsection{Revisiting bandwidth metrics}

Recent literature related to bandwidth measurements in wireless networks has shown how the specific
properties of WLAN access techniques compromise the identifiability of traditional bandwidth
related metrics. In some cases bandwidth metrics require specific definitions
\cite{3,5} to be measured and in some cases new metrics have been defined to account for the particularities of wireless environments \cite{2}.

Traditional metrics associated to bandwidth measurements are \emph{capacity} and \emph{available
bandwidth}. On one side, the capacity, as defined in \cite{1}, is the maximum possible layer-3
(IP) transfer rate at a given network hop or path. As argued above in a wireless environment, this
is a time dependent random process $C(t)$. Even more, the value of the \emph{capacity} depends on
the point of view of the measurement station (e.g. position, transmission rate, quality of channel,
etc.) and strictly depends on the size of the packets used to measure.

On the other side, the \emph{available bandwidth} refers to the portion of the \emph{capacity} that
is not being used: $A(t)$. As happens with the \emph{capacity} metric, it depends on the point of
view of the measurement station and the size of packets being used to measure. Neither the
\emph{available bandwidth}, nor the \emph{capacity} can be normalized for all the nodes that contend
for channel access. As a result, knowing the amount of cross-traffic (in bits-per-second) that
traverses a WLAN link cannot be used to infer the bandwidth available for new stations. Instead,
the impact of cross-traffic has to be measured in terms of the portion of time that the channel is
being used. As a consequence, here we take the definition of \emph{available bandwidth} used in
\cite{3,5} where the authors consider that the \emph{available bandwidth} is the maximum rate that a (measurement) node can transmit without affecting the communication of others.

Recent literature related to bandwidth measurement over wireless systems raised some debate around
the measurement of \emph{available bandwidth} in WLAN scenarios. In \cite{3} the authors show how traditional
techniques fail in measuring such metric in wireless settings. Following such debate and as results
in this paper confirm we propose considering the \emph{achievable throughput} metric in relation to
traditional bandwidth measurement tools. This metric is not new, it was already defined in
\cite{2}. However, the authors proposed an empirical definition of the metric that, as shown here, does not necessarily lead to its actual value. Instead, we propose using the following definition. The rest of the paper uses the term $B$ to refer to the \emph{achievable throughput},

\begin{equation}
\centering \label{eq:IV-I-1} B=sup\{r_i : \frac{r_o}{r_i}\geq1\}
\end{equation}

In this expression, $r_i$ is the rate at which a traffic flow enters the path under measurement and
$r_o$ is the rate at which this traffic leaves the path. Therefore, as considered here, the \emph{achievable throughput} is the maximum rate at which we can inject traffic into a network path and still receive traffic at this same rate. The \emph{achievable throughput} is also a time
varying metric $B(t)$ that depends on the specific characteristics of cross-traffic, channel access
scheduling and channel propagation. Note that under these definitions the relation between the
metrics is $A(t) \leq B(t) \leq C(t)$ (as stated in \cite{2}). This model assumes that, during
the measurement interval, the capacity, the available bandwidth and the achievable throughput are
stationary random processes with asymptotic averages $\overline{C}$,$\overline{A}$ and
$\overline{B}$ respectively.

\subsection{Bandwidth measurements under fluid assumptions}

Here we place fluid assumptions to the WLAN model presented in section 3.1 and review the concept
of the rate response curve \cite{14}. The rate response curve relates the rate ($r_i$) of a
probing flow when it enters a network path with the rate at the output of the path ($r_o$). Fluid
assumptions taken over a wireless link apply to the cross-traffic and to the service rate. Both
processes lose their random and time dependent properties under this assumption and become constant
over the measurement interval. As a consequence, under fluid assumptions we have that during the
entire measurement interval $A(t)=\overline{A}$, $B(t)=\overline{B}$ and $C(t)=\overline{C}$.

Recalling from \cite{14}, the fluid rate response curve of a FIFO queue with constant service rate
(i.e. the probing and cross-traffic share a FIFO queue) can be expressed as,

\begin{equation}
\centering \label{eq:IV-I-2} r_o=min(r_i,C\frac{r_i}{r_i+C-A})
\end{equation}

Note that following our previous discussion we have removed any explicit reference to the amount of
cross-traffic \emph{rate} from this expression (usually called $\lambda$) but we rather express the
equation from the perspective of the node that is measuring.

It is worth noting also that relating equations \ref{eq:IV-I-1} and \ref{eq:IV-I-2} one can see that in a (wired) FIFO system the \emph{achievable throughput}, as defined here, coincides with the \emph{available bandwidth}.

\begin{figure}
\centering
\includegraphics[height=50mm]{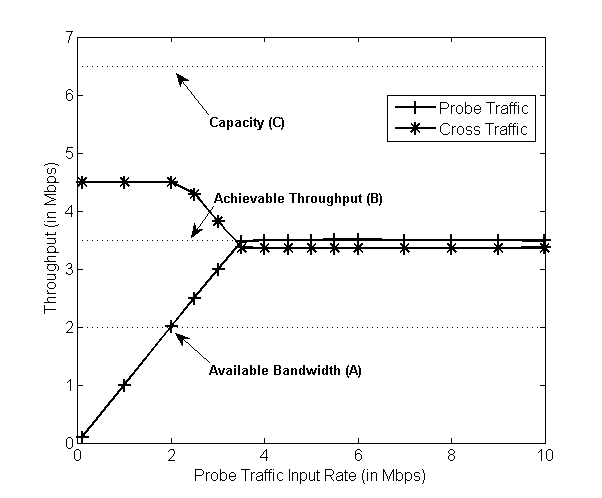}
\caption{Experimental fluid rate response curve of probe traffic in a WLAN setting versus
throughput of cross-traffic flow. C=6.5Mbps, A=2Mbps, B=3.4Mbps} \label{fig:3}
\end{figure}

Now let us consider a case when probing packets contend for (wireless) channel access with the
cross-traffic such as figure \ref{fig:1} depicts. In this case cross-traffic and probing traffic are not scheduled in FIFO order. Figure
\ref{fig:3} plots an experimentation result showing the evolution of the rate response curve when the probing station contends for channel access with another (single) station. In order to obtain the fluid rate response curve we use
long packet probing trains (10000 packets). The figure also shows the evolution of the
cross-traffic throughput for each probing rate. As it can be seen, when the cross-traffic starts
experiencing a decrease in its throughput, that is, when the probing traffic arrives at the
available bandwidth ($\sim$2Mbps), the rate response curve shows no sign of deviation (as one would
expect from eq. \ref{eq:IV-I-2}).  Instead, the rate response curve flattens when the probing rate
reaches the fair share ($\sim$3.5Mbps) that it can get from the wireless medium. This fair share
corresponds, in fact, to the \emph{achievable throughput} metric defined above.

This observation leads to reformulating eq. \ref{eq:IV-I-2} for a wireless link as,

\begin{equation}
\centering \label{eq:IV-I-3} r_o=min(r_i,B)
\end{equation}

Equation \ref{eq:IV-I-3} is the first conclusion of this paper. The fluid rate-response curve of a
wireless link deviates at the \emph{achievable throughput}. Therefore, traditional existing tools
based on this curve \cite{1,17,18,19,20} are, in fact, targeting this metric rather than the \emph{available bandwidth}. Note also that the \emph{available bandwidth} and the \emph{achievable throughput} coincide only when contending stations use a lower portion of the bandwidth than their theoretical fair share.

\section{Non-fluid Analysis of dispersion-based measurements in WLANs}

Recent studies \cite{14,14b,16} have taken a non-fluid approach to the bandwidth measurement problem.
They reveal that dispersion based measurements of the available bandwidth present fundamental deviations from
the fluid model that affect traditional measurement techniques. As they show, such deviations constitute
biases that are difficult to remove when the number of packets used to infer bandwidth metrics (the
train length) is not large enough.

This section takes a similar approach but applied to bandwidth measurements in WLAN environments.

\subsection{Analytical framework}

Here we present the basic analytical framework used to deal with this problem. This framework was originally proposed in \cite{7} but is extended here to focus on the particularities of WLAN transmissions.

\subsubsection{The probing sequence: Arrivals, departures and input gap}

The probing sequence consists of a series of $n$ packets that enter the transmission queue at
instants $\{a_i,i=1,2,\cdots,n\}$. Their departure instants, meaning the time at which they are
completely transmitted, form the series $\{d_i,i=1,2,\cdots,n\}$. Finally, we are considering here
periodic probing flows with a fixed inter-packet arrival time or input gap: $g_I=a_i-a_{i-1}$.

\subsubsection{The service delay process}

As shown above, the service delay that probing packets experience is a random process. This process
is the result of the interaction between probing traffic, contending cross-traffic and backoff. To
account for this let us define the sequence $\{\mu_i,i=1,2,\cdots,n\}$ to denote the random service
delay that each one of the $n$ probing packets of a probing sequence experiences when contending
for medium access.

As shown above, the service delay presents a transitory period until reaching a certain stationary
distribution.  Thus, $\exists n_0: \forall \{i>n_o\}, \mu_i$ is i.i.d. Further, we assume that the
service delay distribution is upper and lower bounded. In other words, we assume that  $\exists
\{\mu^{max},\mu^{min}\}:\forall i, Pr(\mu^{min} \leq \mu_i \leq \mu^{max})=1$.

\subsubsection{Intrusion residual: amount of probe traffic in the FIFO queue}

The intrusion residual $W_d(t)$ accounts for the sum of the service time of all probing packets in
the FIFO queue and the remaining time to service the probing packet that may be in transmission.
Next, we define the series  $\{R_i,i=1,2,\cdots,n\}$ which captures the intrusion residual that
every probing packet finds when it enters the transmission queue,

\begin{equation}
\centering \label{eq:IV-I-4} R_i(a_1)=W_d(a_i^-)=W_d(a_1+(i-1)g_I^-)
\end{equation}

Note\footnote{The minus superscript refers to the \emph{a priori} state of the queue.} that $R_i$
is a recursive process that under the assumptions in this work can be expressed as,

\begin{equation}
R_i =
\begin{cases}
         0 & i=1 \\
         max(0,\mu_{i-1}+R_{i-1}-g_I) & i>1
         \end{cases}
\label{eq:IV-I-5}
\end{equation}

Finally, we define the series $\{Z_i,i=1,2,\cdots,n\}$ that encloses the queuing plus service delay
that each one of the probing packets experiences. Under the assumptions taken,

\begin{equation}
\centering \label{eq:IV-I-6} Z_i=d_i-a_i=\mu_i+R_i
\end{equation}

\subsubsection{Dispersion based measurements:The output gap and its relation to the probing rate}

Dispersion based measurements of bandwidth metrics consist on measuring the dispersion (or
inter-departure time) of packets at the output of a path (receiving side). This measure is then
used to infer the value of bandwidth related metrics. The output gap (or dispersion) of a train of
probing packets is defined as follows,

\begin{equation}
\centering \label{eq:IV-I-7} g_O=\frac{d_n-d_1}{n-1}
\end{equation}

\begin{figure}
\centering
\includegraphics[height=30mm]{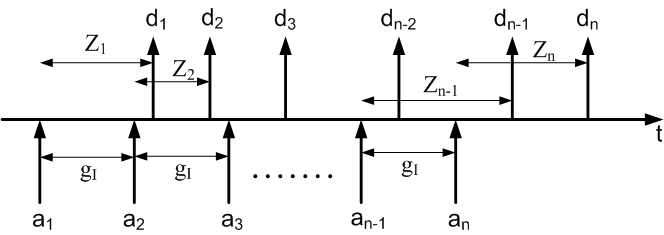}
\caption{Inter-relation between probing arrival sequence ($a_i$), departure sequence ($d_i$) and
cross-traffic related processes ($Z_i$).} \label{fig:11}
\end{figure}

Figure \ref{fig:11} illustrates the contribution of the processes defined above to the value of the
output gap. From the arrival of the first probing packet at the transmission queue ($a_1$), probing
packets keep on arriving at a constant interval of $g_I$. The cross-traffic, service delay and the
intrusion residual of previous probing packets ($Z_i$) randomize the departure times of probing
packets ($d_i$) and thus, their output dispersion ($g_O$).

Observing figure \ref{fig:11} we can obtain the output gap in relation to the different processes
involved.

\begin{equation}
\centering \label{eq:IV-I-8} g_O=\frac{d_n-d_1}{n-1}=\frac{(n-1)g_I + Z_n - Z_1}{n-1}
\end{equation}

Expanding this expression we get the following,

\begin{equation}
\centering \label{eq:IV-I-9} g_O=g_I+\frac{R_n}{n-1}+\frac{\mu_n-\mu_1}{n-1}
\end{equation}

\subsubsection{Problem formulation}

We are interesting in studying whether dispersion measurements can be used to infer the rate
response curve of a wireless link. Measurement tools based on dispersion take the assumption that
the relation between the input ($g_I$) and output ($g_O$) dispersions of a probing train can be
used as estimators of the inter-relation between input ($r_i$) and output ($r_O$) rates of a flow
traversing the system. In other words if $L$ is the length of the packets used for probing,
dispersion based measurements assume that $L$/$g_I$ is equivalent to $r_i$ and the same between
$L$/$g_O$ and $r_O$.

Reformulating equation \ref{eq:IV-I-3}, the problem of bandwidth estimation using dispersion
measurements can be stated as follows.

\begin{equation}
E[g_O] \stackrel{?}{=}
\begin{cases}
         g_I & g_I \geq \frac{L}{\overline{B}} \\
         \frac{L}{\overline{B}} & g_I \leq \frac{L}{\overline{B}}
         \end{cases}
\label{eq:IV-I-11}
\end{equation}

Taking expectation over eq. \ref{eq:IV-I-9}, the rest of this section deals with the evaluation of
the behavior of the following expression,

\begin{equation}
\centering \label{eq:IV-I-12} E[g_O]=g_I+\frac{E[R_n]}{n-1}+\frac{E[\mu_n]-E[\mu_1]}{n-1}
\end{equation}

\subsection{The impact of the randomness of service delay on dispersion measurements}

This section studies the impact on dispersion measurements of the randomness of the service delay.
Along this section we assume that the service delay does not present the transitory stage described
in section 3.2. Instead we assume that, for all probing packets, the service delay is i.i.d. This
assumption would apply in a WLAN link with no contending traffic but in which the quality of the
channel leads to frequent retransmissions or frequent changes of the transmission rate.

The analysis reveals how dispersion measurements, when taken around the \emph{achievable
throughput} present a deviation from the fluid response curve. Interestingly enough the origin of
this deviation is similar to the one detected in \cite{14}. This indicates that the randomness of
the service delay causes a similar effect as the burstiness of cross-traffic in a (wired) FIFO
queue.

\subsubsection{Expected output dispersion and achievable throughput}

Assuming that the service delay does not present the transitory stage detected in section 3.2,
expression \ref{eq:IV-I-12} reduces to the following,

\begin{equation}
\centering \label{eq:IV-I-13} E[g_O]=g_I+\frac{E[R_n]}{n-1}
\end{equation}

Under the assumption of no other (cross-)traffic in the FIFO queue the system can serve, in
average, up to one probing packet every $E[\mu]$. As a consequence we can state that,\\

\begin{equation}
\centering \label{eq:IV-I-14} \overline{B}=\frac{L}{E[\mu]}
\end{equation}\\
\\

\subsubsection{Expectation on output gap based on bounds of the intrusion residual}

From expression \ref{eq:IV-I-13}, we learn that the expected output gap depends on the expected
value for the residual that the last packet of the probing train (i.e. with index $n$) finds in the
queue. Considering eq. \ref{eq:IV-I-5} and that the service delay presents upper and lower bounds
(i.e. $\mu^{max}$ and $\mu^{min}$), one can define the following (loose) bounds for the probing
residual:

\begin{equation}\small
\begin{cases}
         R_n = \sum_{i=1}^{n-1}(\mu_i-g_I) & g_I \leq \mu^{min} \\
         max(0,\sum_{i=1}^{n-1}(\mu_i-g_I)) \leq R_n \leq \sum_{i=1}^{n-1}\mu_i & \mu^{min} \leq g_I \leq \mu^{max} \\
         R_n = 0 & g_I \geq \mu^{max}
\end{cases}
\label{eq:IV-I-15}
\end{equation}

Taking expectation over $R_n$, we can identify four differentiated regions,

\begin{equation}
\frac{E[R_n]}{n-1} =
\begin{cases}
E[\mu] - g_I & g_I \leq \mu_s^{min} \\
\frac{\beta_n}{n-1} & \mu^{min} \leq g_I \leq E[\mu] \\
\frac{\alpha_n}{n-1} & E[\mu] \leq g_I \leq \mu^{max} \\
0 & g_I \geq \mu^{max}
\end{cases}
\label{eq:IV-I-16}
\end{equation}

The parameters $\alpha_n$ and $\beta_n$ depend on the specific characteristics of the random
cross-traffic but can be (loosely) bounded as follows,

\begin{equation}
\begin{cases}
    E[\mu]-g_I \leq \frac{\beta_n}{n-1} \leq E[\mu] \\
    0 \leq \frac{\alpha_n}{n-1} \leq E[\mu]
\end{cases}
\label{eq:IV-I-17}
\end{equation}

Finally, we are interested in the output dispersion. Thus, substituting eq. \ref{eq:IV-I-16} into
eq. \ref{eq:IV-I-13} we get the following,

\begin{equation}
E[g_O] =
\begin{cases}
      E[\mu] & g_I \leq \mu^{min} \\
      g_I + \frac{\beta_n}{n-1} & \mu^{min} \leq g_I \leq E[\mu] \\
      g_I + \frac{\alpha_n}{n-1} & E[\mu] \leq g_I \leq \mu^{max} \\
      g_I & g_I \geq \mu^{max}
\end{cases}
\label{eq:IV-I-18}
\end{equation}

Two important observations about eq. \ref{eq:IV-I-18} are that, first, $\alpha_n$ and $\beta_n$ are
deviation terms that depend on the length of the packet train ($n$) and that disappear when the
input gap ($g_I$) falls outside the limits of the random service delay. Second, the lower bound of
the output gap corresponds to the fluid response curve.

\subsubsection{Numerical results on the rate response curve}

\begin{figure}
\centering
\includegraphics[height=50mm]{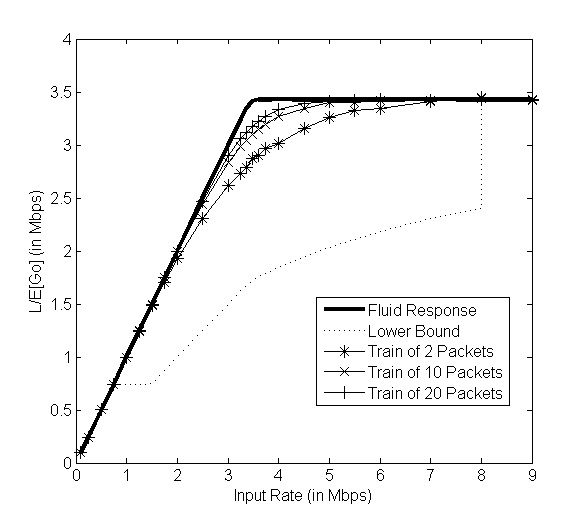}
\caption{Rate response curve when probing a system with exponentially distributed service delay
with probing sequences of different length.} \label{fig:13}
\end{figure}

In order to obtain numerical evidence of the expressions presented above we have used the queuing
simulator introduced in section 2. Figure \ref{fig:13} plots the expected rate response curve
inferred using the dispersion of probing trains of different lengths (2, 10 and 20 packets) at the
output of a system with exponentially distributed service delay (see \cite{7}). The mean service
delay that probing packets experience corresponds to an achievable throughput of 3.5Mbps (probing
packets are 1500 bytes long). The fluid response curve is also included. The figure shows how when
probing around the \emph{achievable throughput} dispersion measurements deviate from the fluid
response curve. The deviation is higher the lower the number of packets used to measure. The figure
plots also, as a reference, the (lower) bound on the maximum deviation of the rate response curve
coming from the (upper) bound on the expected output gap derived in expression \ref{eq:IV-I-18}.

\subsubsection{The origin of the bias}

The biases detected in eq. \ref{eq:IV-I-18} and shown in figure \ref{fig:13} have their origin in
the evolution of the expected queuing delay that probing packets suffer when traversing the
transmission queue before being served (see \ref{fig:2} for a reference). To illustrate this,
figure \ref{fig:14} plots, for various probing rates, the difference in the mean delay experienced
by each one of the first 100 packets of a long probing flow with respect to their immediate
predecessor. In other words it plots the process $\{E[Z_i-Z_{i-1}],i=2,\cdots,n\}$. It can be seen
from the figure that it takes some probing packets until the process $E[Z_i-Z_{i-1}]$ becomes
stable (constant). This is precisely what deviates the $\alpha_n$ and $\beta_n$ terms in equation
\ref{eq:IV-I-17} from the fluid response curve.

This figure reveals that, when probing packets are served with random delay, it takes some packets
until they start experiencing a stationary behavior. The first packets, then, constitute biased
measures of the delay required to traverse the link. In other words, first packets are not able to
capture the stationary behavior of the queue state and distort dispersion measurements.

\subsubsection{Asymptotics of the deviation terms}

\begin{figure}
\centering
\includegraphics[height=50mm]{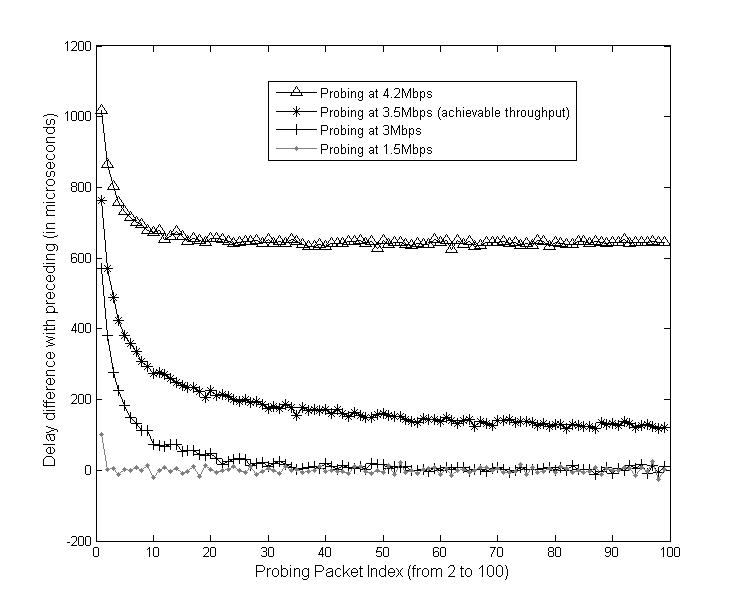}
\caption{Expected per-packet delay difference between consecutive probing packets sent through a
FIFO queue with exponential service delay} \label{fig:14}
\end{figure}

As figure \ref{fig:14} suggests, the longer the packet train the lower the bias of dispersion
measurements. It can be shown that eq. \ref{eq:IV-I-18} tends asymptotically to the fluid response
curve as $n$ increases. That is,

\begin{equation}
\centering \label{eq:IV-I-19} \lim_{n\rightarrow+\infty}\frac{\alpha_n}{n-1}=0
\end{equation}

and,

\begin{equation}
\centering \label{eq:IV-I-20} \lim_{n\rightarrow +\infty}\frac{\beta_n}{n-1}=E[\mu]-g_I
\end{equation}

Furthermore, if we unbound the service delay, it can be shown that there exists a strong relation
between the variance of the service delay and the intensity of the deviation of dispersion
measurements from the rate response curve. We have that,

\begin{equation}
\centering \label{eq:IV-I-21} \lim_{Var[\mu]\rightarrow +\infty}\frac{\alpha_n}{n-1}=E[\mu]
\end{equation}

and,

\begin{equation}
\centering \label{eq:IV-I-22} \lim_{Var[\mu]\rightarrow +\infty}\frac{\beta_n}{n-1}=E[\mu]
\end{equation}

\subsubsection{A final remark}
The findings in this section present high similarities with the ones presented in \cite{14}. This
indicates that the impact on probing traffic of a random service delay process (i.i.d) is similar
to the case when probing packets share a (wired) FIFO queue with a bursty cross-traffic flow.

\subsection{The impact of the transitory regime of service delay on dispersion measurements}

This section reintroduces the transitory of the service delay and studies its impact on dispersion
measurements.

The basic finding here is that the transitory of the service delay induces new deviations of the
rate response curve, but in the opposite direction than the pure i.i.d random service. As a
consequence packets in short probing trains are 'accelerated' in comparison to packets in longer
trains. This may lead to obtaining overestimates of the actual rate response curve when using
dispersion measurements.

\subsubsection{Expected output dispersion and achievable throughput}

Now the expression of the output gap cannot be reduced and our objective is studying this
expression,

\begin{equation}
\centering \label{eq:IV-I-23} E[g_O]=g_I+\frac{E[R_n]}{n-1}+\frac{E[\mu_n]-E[\mu_1]}{n-1}
\end{equation}

We can define again a relation between the achievable throughput and the service delay that probing
packets receive.

\begin{equation}
\centering \label{eq:IV-I-24} \frac{L}{\overline{B}}=\frac{1}{n}\sum_{i=1}^n(E[\mu_i])
\end{equation}

Note also that as the number of probing packets grows the expected service delay becomes constant
and we can say that,

\begin{equation}
\centering \label{eq:IV-I-25} \frac{L}{\overline{B}} \stackrel{n}{\rightarrow} E[\mu_n]
\end{equation}

\subsubsection{Expectation on output gap based on bounds of the intrusion residual}

Following similar reasoning as in the previous section, the expected output dispersion of a train
of $n$ packets presents four differentiated regions such as,

\begin{equation}
E[g_O] =
\begin{cases}
    \frac{1}{n-1}(\sum_{i=2}^{n}({E[\mu_i]})) & g_I \leq \mu^{min} \\
      g_I + \frac{\beta_n}{n-1} & \mu^{min} \leq g_I \leq \frac{1}{n}\sum_{i=1}^{n}{E[\mu_i]} \\
      g_I + \frac{\alpha_n}{n-1} & \frac{1}{n}\sum_{i=1}^{n}{E[\mu_i]} \leq g_I \leq \mu^{max} \\
      g_I + \frac{[\mu_n-\mu_1]}{n-1} & g_I \geq \mu^{max} \\
\end{cases}
\label{eq:IV-I-26}
\end{equation}

The parameters $\alpha_n$ and $\beta_n$ in the expression above are bounded as follows,

\begin{equation}
\begin{cases}
       \frac{1}{n}\sum_{i=2}^{n}({E[\mu_i]-g_I}) \leq \frac{\beta_n}{n-1} \leq \frac{1}{n-1}\sum_{i=2}^{n}({E[\mu_i]}) \\
       0 \leq \frac{\alpha_n}{n-1} \leq \frac{1}{n-1}\sum_{i=2}^{n}({E[\mu_i]})
\end{cases}
\label{eq:IV-I-27}
\end{equation}

Expressions \ref{eq:IV-I-26} and \ref{eq:IV-I-27} reveal some interesting features about dispersion
measurements for WLAN environments.

First, note that considering that the expected service delay in a WLAN environment is an increasing
function with respect to the packet index ($i$), the following is true for any value of $n$,

\begin{equation}
\centering \label{eq:IV-I-28} \frac{1}{n-1}\sum_{i=2}^{n}({E[\mu_i]}) < E[\mu_n]
\end{equation}

As a result, when probing packets arrive faster than the achievable throughput (i.e.
$g_I\leq\frac{1}{n}\sum_1^nE[\mu_i])$, packets at the output experience a 'compression' effect with
respect to the fluid response that leads to inferring higher output rates than those that can be
actually achieved in this region.

Second, as eq. \ref{eq:IV-I-27} reveals, the upper bound of $\alpha_n$ is lower than without the
presence of the transitory. The bias introduced by this term in dispersion measurements is, thus,
lower than would be expected without the presence of the transitory.

Third, when the input rate is low enough (i.e. when $g_I\geq\mu^{max}$) the transitory causes an
expansion of the expected output dispersion. This effect is hidden in rate response curves

\begin{figure}
\centering
\includegraphics[height=50mm]{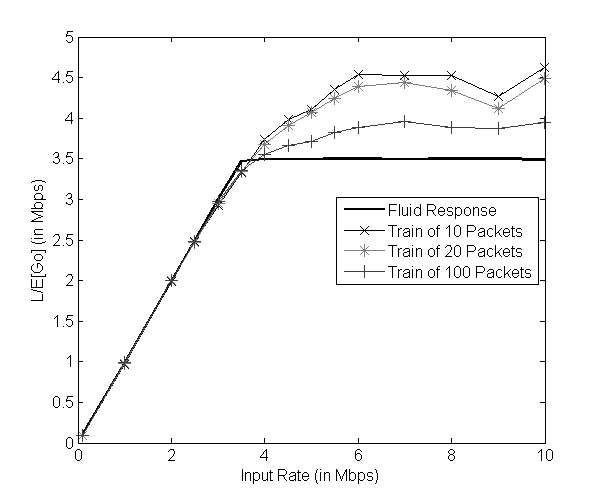}
\caption{Experimental rate response curve when probing a WLAN system with probing sequences of
different length} \label{fig:15}
\end{figure}

Figure \ref{fig:15} shows an experimental result showing these observations. The rate response
curves plotted correspond to those of packet trains probing a WLAN link at different rates. The
figure clearly shows how, when short packet trains are used, the rate response curve gathered leads
to inferring higher rates than the achievable throughput. The figure also shows how the bias
introduced by the $\alpha_n$ term and the expansion suffered at low probing rates are not important
enough to distort the measurement process.

\section{Discussion on consequences and applications of findings}

This section discusses the main findings of this study and some consequences and possible applications that they entail.

\subsection{Summary of findings}

\begin{itemize}
\item{In section 4 we showed how the rate response curve, when applied over a WLAN system,
presents a deviation point at the \emph{achievable throughput} rather than the \emph{available
bandwidth}}
\item{In section 5 we showed how dispersion based measurements are biased estimations of the rate response curve.
There are two possible sources of bias in a WLAN system. On one side there is the randomness of the
service delay that probing packets may experience. On the other side, there is the transitory in
distribution that the random service delay presents when probing traffic contents for channel
access. In both cases the origin of the bias lies in the fact that it takes a while (some packets)
for the probing traffic to completely interact with the system. This implies that the first packets
in a probing sequence are not valid samples of the stationary behavior of the system.}
\end{itemize}

\subsection{A consequence: bandwidth estimation in WLAN links}

As shown in section 4, traditional methodologies developed taking the rate response curve as a reference target the \emph{achievable throughput} rather than the \emph{available bandwidth}. The same happens with those tools designed to infer the \emph{capacity} based on dispersion measurements.

At this point arises the question of whether it is worth measuring the \emph{achievable throughput} of a network path. We argue that such a metric can be useful for routing protocols, overlay/P2P network formations or even in congestion control algorithms as it allows knowing the throughput that a given node will receive when sending data without distortion. As an example, for TCP congestion control, it would allow maintaining a more efficient control of the evolution of the congestion window during the congestion avoidance phase. However, the \emph{available bandwidth} metric is still of much interest for applications such as access control algorithms or to tune the slow start phase of TCP congestion control. As shown in section 4 however, the rate response curve does not present any signal at the available bandwidth and novel methodologies such as the one proposed in \cite{3} should be developed.

\begin{figure}
\centering
\includegraphics[height=50mm]{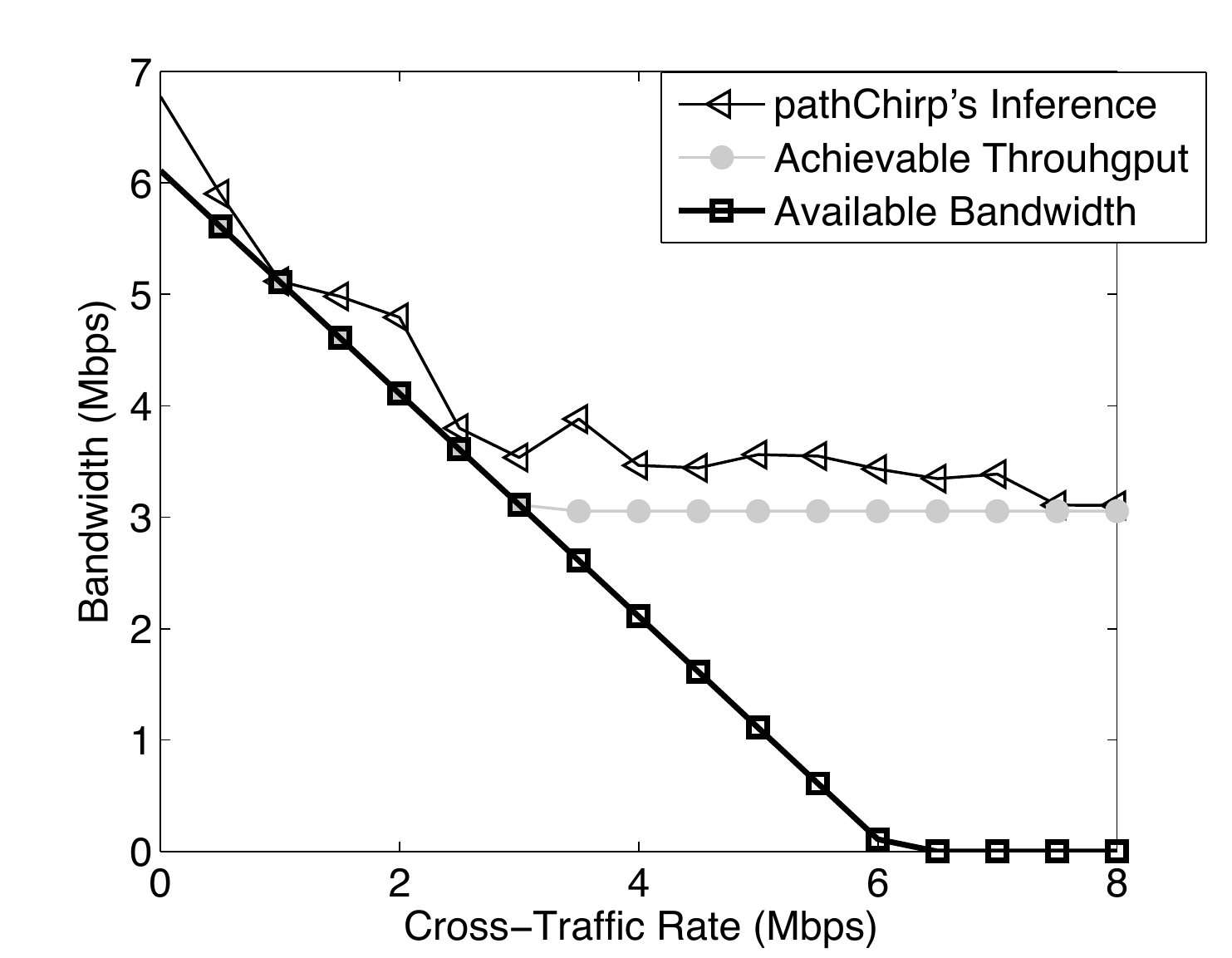}
\caption{Estimation of pathChirp in a wireless link (1 contending node, exponential inter-departure
time, 1500bytes as packet size, intensity varies). The tool follows the achievable throughput rather than the available bandwidth.} \label{fig:16}
\end{figure}

To illustrate this consider the experiment depicted in figure \ref{fig:16} (NS2). We have run a state-of-the-art
available banwdith estimation tool (pathChirp \cite{19}) in the presence of a wireless link. This tool is designed with the rate response curve as a reference and tries to find a turning point in the curve using dispersion of packets. As the figure shows pathChirp points at the \emph{achievable throughput}. This becomes clear when the available bandwidth and achievable throughput deviate one from the other (at around 3Mbps in this case).

\subsection{Another consequence: packet pair measurements in WLAN links}

A common approach to measure the capacity of a network path is the packet-pair technique\cite{24}. Recently, packet pairs have gained momentum as they have been extensively used to develop routing metrics in all-wireless multi-hop networks \cite{22}.

\begin{figure}
\centering
\includegraphics[height=50mm]{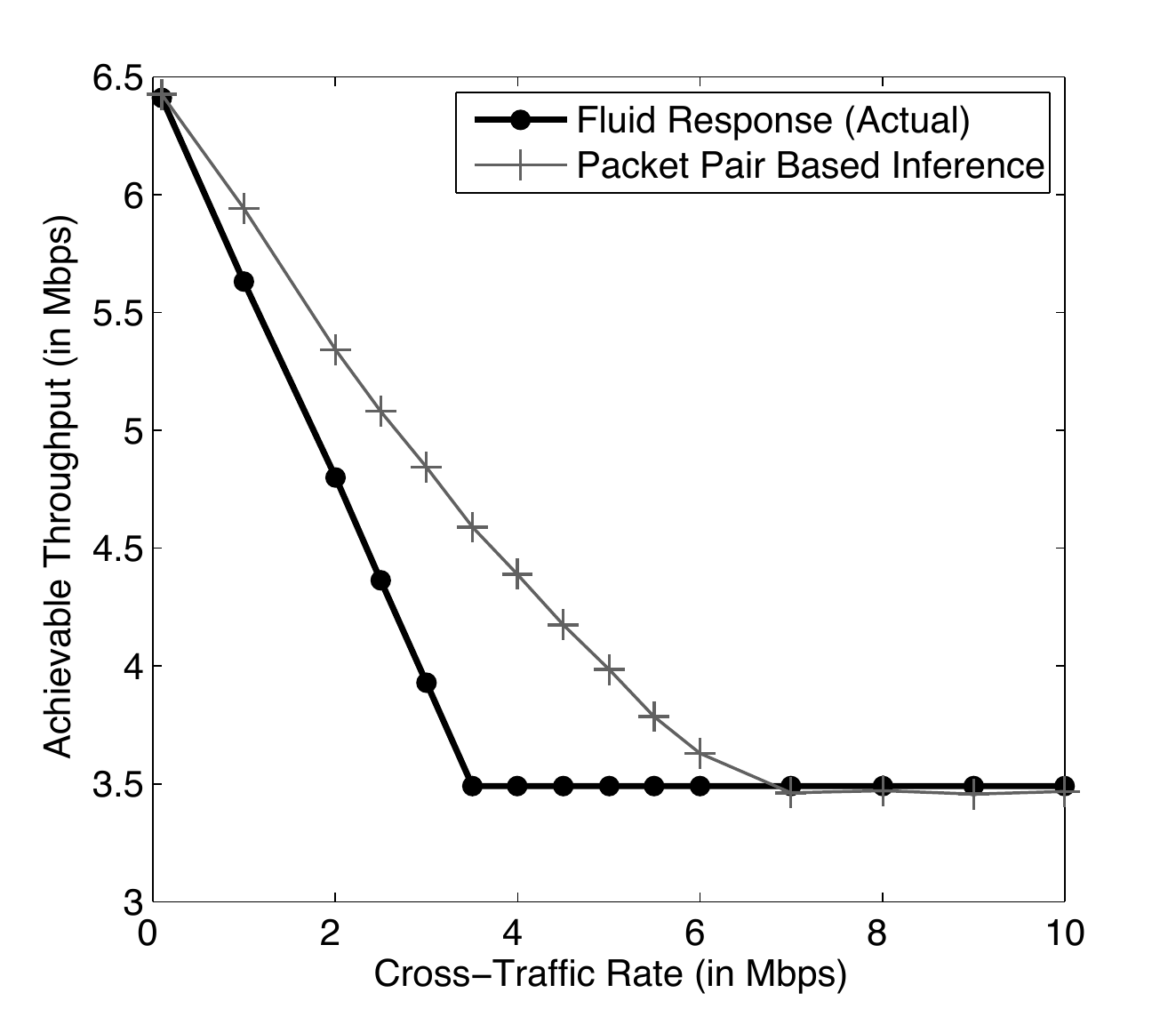}
\caption{Experimental comparison between packet pair based bandwidth measurements and the actual
fluid response in a WLAN link} \label{fig:17}
\end{figure}

However, as a consequence of the results presented in section 4, packet pairs (understood as probes of infinite rate) target the \emph{achievable throughput} when used in a WLAN link. Even more, considering results from section 5.3, one can see that packet pairs tend to overestimate the value of the \emph{achievable throughput}. Figure \ref{fig:17} illustrates this fact. It plots the actual \emph{achievable throughput} of a WLAN link and the estimation using average dispersion of packet-pairs. This is done for different levels of cross-traffic. The \emph{capacity} of the WLAN link is kept constant for all the measurement process at 6.5Mbps (no channel propagation errors). As one can see the packet-pair does not measure the \emph{capacity} in the whole measurement region except when no contending traffic is present.

\subsection{An application of results: improving the convergence and accuracy of traditional tools}

The results presented in section 5 entail a second important observation that can be used to
improve the accuracy of bandwidth measurement tools. As figure \ref{fig:4} and \ref{fig:14}
reveal, the first packets that traverse a WLAN link constitute non-accurate samples of the
stationary behavior. A direct implication of this is that first samples (packets) of the probing
train should be removed from bandwidth estimates as they are not accurate.

Traditionally the approach to remove measurement biases consists in enlarging the number of packets
used to gather measurements. However, this comes at the cost of increasing the intrusiveness of the
measurement process over the measured path. The results in section 5 lead to the observation that removing the first
packet samples from bandwidth measurements helps reducing the measurement bias and can help improve
the measurement accuracy with a limited number of probing packets.

\begin{figure}
\centering
\includegraphics[height=50mm]{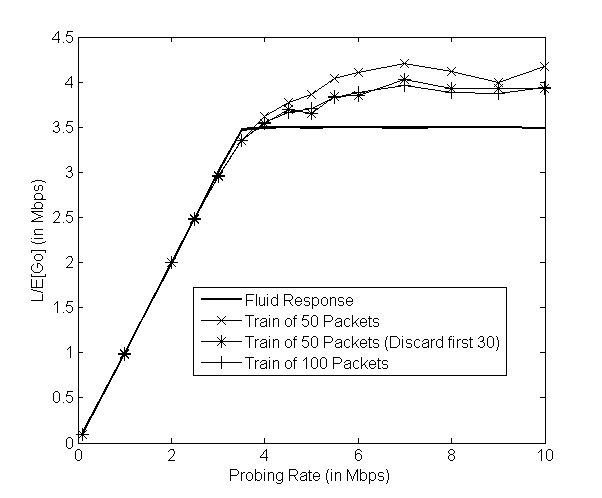}
\caption{Experimental rate response curves showing the bias incurred by different probing
strategies} \label{fig:18}
\end{figure}

Figure \ref{fig:18} illustrates this observation. As the figure shows one can achieve the same
measurement accuracy using trains of 50 packets (but removing the first 30 from the measure) as
when using trains of 100 packets. This could be easily applied to existing tools
\cite{1,17,18,19,20,23,24} improving their accuracy and/or reducing their convergence time.

Interestingly enough, the similarity of results in section 5.2 and the results presented in \cite{14} suggests that removing the first packets from measurements of \emph{available bandwidth} taken over wired paths would also increase the accuracy of the result.

\section{Conclusions}

This paper presents a study of the bandwidth measurement problem over WLANs. The paper models the time to service probing packets as a random process and analyzes the interaction between periodic probing sequences and this random process. The approach is non-parametric.

The analysis reveals how the randomness of the WLAN service time introduces a bias in measurement techniques based on the dispersion of probing packets. Interestingly, this bias presents strong similarities with the one reported in \cite{14,14b} which suggests that bursty cross-traffic can be assimilated as any other random process affecting a periodic probing sequence. Hence, the results presented here can be made extensible to any other system containing any form of randomness in the time to transmit packets (e.g. PLC links). Further, for the particular case of an IEEE 802.11 link, the paper reveals that the service delay does not follow a stationary process but, instead, presents a transitory regime. This transitory introduces additional bias in bandwidth measurements which lead to an overestimation of the rate-response curve of a WLAN link when probing at high packet rates.

Several important problems remain open for future work. First this analysis, as well as \cite{14,14b}, focuses on periodic probing sequences. Extending this analysis to other probing processes such as Poisson may lead to interesting conclusions. Alternative probing sequences present also the same transitory  behavior as periodic probing sequences and may also be subject to similar biases. Second, we have seen that dispersion measurements are not useful to measure the \emph{available bandwidth}. Therefore, it would be interesting to explore whether other mechanisms can be used for this purpose.

\bibliographystyle{unsrt}

\end{document}